\input prepictex
\input pictex
\input postpictex
\newcommand{\ptcommand}[1]{\put(0,0){\beginpicture
\setcoordinatesystem units <\unitlength,\unitlength>
#1
\endpicture}}
\documentstyle[12pt]{article}
\renewcommand{\baselinestretch}{1.5}
\renewcommand{\arraystretch}{1.5}
\begin{document}
\thispagestyle{empty}
\pagestyle{empty}
\renewcommand{\thefootnote}{\fnsymbol{footnote}}
\newcommand{\preprint}[1]{\begin{flushright}
\setlength{\baselineskip}{3ex}#1\end{flushright}}
\renewcommand{\title}[1]{\begin{center}\LARGE
#1\end{center}\par}
\renewcommand{\title}[1]{\begin{center}\LARGE #1\end{center}\par}
\renewcommand{\author}[1]{\vspace{2ex}{\Large\begin{center}
\setlength{\baselineskip}{3ex}#1\par\end{center}}}
\renewcommand{\thanks}[1]{\footnote{#1}}
\renewcommand{\abstract}[1]{\vspace{2ex}\normalsize\begin{center}
\centerline{\bf Abstract}\par\vspace{2ex}\parbox{6in}{#1
\setlength{\baselineskip}{2.5ex}\par}
\end{center}}
\newcommand{\starttext}{\newpage\normalsize
\pagestyle{plain}
\setlength{\baselineskip}{4ex}\par
\setcounter{footnote}{0}
\renewcommand{\thefootnote}{\arabic{footnote}}
}

\newcommand{\segment}[2]{\put#1{\circle*{2}}}
\newcommand{\fig}[1]{figure~\ref{#1}}
\newcommand{\hc}{{\rm h.c.}}
\newcommand{\ds}{\displaystyle}
\newcommand{\eqr}[1]{(\ref{#1})}
\newcommand{\tr}{\,{\rm tr}}
\newcommand{\uone}{{U(1)}}
\newcommand{\su}[1]{{SU(#1)}}
\newcommand{\stu}{\su2\times\uone}
\newcommand{\be}{\begin{equation}}
\newcommand{\ee}{\end{equation}}
\newcommand{\bp}{\begin{picture}}
\newcommand{\ep}{\end{picture}}
\def\spur#1{\mathord{\not\mathrel{#1}}}
\newcommand{\sechead}[1]{\medskip{\bf #1}\par\bigskip}
\newcommand{\ba}[1]{\begin{array}{#1}\ds }
\newcommand{\cra}{\\ \ds}
\newcommand{\ea}{\end{array}}
\newcommand{\forto}[3]{\;{\rm for}\; #1 = #2 \;{\rm to}\; #3}
\newcommand{\for}{\;{\rm for}\;}
\newcommand{\cross }{\hbox{$\times$}}
\newcommand{\ol}{\overline}
\newcommand{\bra}[1]{\left\langle #1 \right|}
\newcommand{\ket}[1]{\left| #1 \right\rangle}
\newcommand{\braket}[2]{\left\langle #1 \left|#2\right\rangle\right.}
\newcommand{\braketr}[2]{\left.\left\langle #1 right|#2\right\rangle}
\newcommand{\g}[1]{\gamma_{#1}}
\newcommand{\half}{{1\over 2}}
\newcommand{\del}{\partial}
\newcommand{\grad}{\vec\del}
\newcommand{\real}{{\rm Re\,}}
\newcommand{\imag}{{\rm Im\,}}
\newcommand{\gapprox}{\raisebox{-.2ex}{$\stackrel{\textstyle>}
{\raisebox{-.6ex}[0ex][0ex]{$\sim$}}$}}
\newcommand{\lapprox}{\raisebox{-.2ex}{$\stackrel{\textstyle<}
{\raisebox{-.6ex}[0ex][0ex]{$\sim$}}$}}
\newcommand{\cl}[1]{\begin{center} #1\end{center}}
\newcommand\etal{{\it et al.}}
\newcommand{\prl}[3]{Phys. Rev. Letters {\bf #1} (#2) #3}
\newcommand{\prd}[3]{Phys. Rev. {\bf D#1} (#2) #3}
\newcommand{\npb}[3]{Nucl. Phys. {\bf B#1} (#2) #3}
\newcommand{\plb}[3]{Phys. Lett. {\bf #1B} (#2) #3}
\newcommand{\ie}{{\it i.e.}}
\newcommand{\etc}{{\it etc.\/}}
\def\cA{{\cal A}}
\def\cB{{\cal B}}
\def\cC{{\cal C}}
\def\cD{{\cal D}}
\def\cE{{\cal E}}
\def\cF{{\cal F}}
\def\cG{{\cal G}}
\def\cH{{\cal H}}
\def\cI{{\cal I}}
\def\cJ{{\cal J}}
\def\cK{{\cal K}}
\def\cL{{\cal L}}
\def\cM{{\cal M}}
\def\cN{{\cal N}}
\def\cO{{\cal O}}
\def\cP{{\cal P}}
\def\cQ{{\cal Q}}
\def\cR{{\cal R}}
\def\cS{{\cal S}}
\def\cT{{\cal T}}
\def\cU{{\cal U}}
\def\cV{{\cal V}}
\def\cW{{\cal W}}
\def\cX{{\cal X}}
\def\cY{{\cal Y}}
\def\cZ{{\cal Z}}
\renewcommand{\baselinestretch}{1.5}
\renewcommand{\arraystretch}{1.5}
\newcommand{\boxit}[1]{\ba{|c|}\hline #1 \\ \hline\ea}
\newcommand{\mini}[1]{\begin{minipage}[t]{20em}{#1}\vspace{.5em}
\end{minipage}}

\def\theequation{\themysection.\arabic{equation}}
\def\eps{\epsilon}
\def\pre{{- i \mu^\eps \over 16 \pi^2 \eps}}
\def\ide{ \int d \Omega^4_e \,}
\def\pref{{- i \mu^\eps g^2 \over 32 \pi^4 \eps}}
\def\bea{\begin{eqnarray}}
\def\eea{\end{eqnarray}}

\preprint{\#HUTP-92/A060\ \\ 5/92 }
\title{Running Nonlocal Lagrangians\thanks{Research
supported in part by the National Science Foundation under Grant
\#PHY-8714654.}\thanks{Research supported in part by the Texas National
Research Laboratory Commission, under Grant \#RGFY9106.}
}
\author{Vineer Bhansali and Howard Georgi \\
{\large Lyman Laboratory of Physics,
Harvard University \\
Cambridge, MA 02138}
}
\date{}

\abstract{
We investigate the renormalization of ``nonlocal'' interactions in an
effective field theory using dimensional regularization with minimal
subtraction. In a scalar field theory, we write an integro-differential
renormalization group equation for every possible class of graph at one loop
order.}

\starttext

\def\theequation{\arabic{equation}}

\section{ Introduction}

In its traditional form, an effective field theory calculation goes like this:
Start at a very large scale, that is with the renormalization scale, $\mu$,
very large. In a strongly interacting theory or a theory with unknown physics
at high energy, this starting scale should be sufficiently large that
nonrenormalizable interactions produced at higher scales are too small to be
relevant. In a renormalizable, weakly interacting theory, one starts at a
scale above the masses of all the particles, where the effective theory is
given simply by the renormalizable theory, with no nonrenormalizable terms.
The theory is then evolved down to lower scales. As long as no particle masses
are encountered, this evolution is described by the renormalization group.
However, when $\mu$ goes below the mass, $\Lambda$, of one of the particles in
the theory, we must change the effective theory to a new theory without that
particle. In the process, the parameters of the theory change, and new,
nonrenormalizable interactions may be introduced. Both the changes in existing
parameters, and the coefficients of the new interactions are computed by
``matching'' the physics just below the boundary in the two theories. It is
this process that determines the relative sizes of the nonrenormalizable terms
associated with the heavy particles.

Because matching is done for $\mu\approx\Lambda$, the rule for the size of the
coefficients of the new operators is simple for $\mu\approx\Lambda$. At this
scale, all the new contributions scale with $\Lambda$ to the appropriate power
(set by dimensional analysis) up to factors of coupling constants, group
theory or counting factors and loop factors (of $16\pi^2$, \etc) \cite{GM}.
Then when the new effective theory is evolved down to smaller $\mu$, the
renormalization group introduces additional factors into the coefficients.
Thus a heavy particle mass appears in the parameters of an effective field
theory in two ways. There is power dependence on the mass that arises from
matching conditions. There is also logarithmic dependence that arises from the
renormalization group.

The matching correction at tree level is simply a difference between a
calculation in the full theory and a calculation in the low energy effective
theory
\be\ba{c}
\int\delta\cL^0(\Phi)=S_{\cL_H+\cL}^0(\Phi)-S^0_{\cL}(\Phi)\cra
=\int{\textstyle\left\{
{\rm virtual\;heavy\atop particle\;trees}\right\}}(\Phi)\ea
\ee
where $S_{\cL_H+\cL}^0(\Phi)$ denotes the light particle effective action in
the full theory and $S^0_{\cL}(\Phi)$ denotes the same in the low energy
theory \cite{HG}.
The matching correction so obtained is nonlocal because it depends on
$p/\Lambda$ through the virtual heavy particle propagators. It is also \bf
analytic \rm in $p/\Lambda$ in the region relevant to the low energy theory,
i.e. for characteristic momentum $<< \Lambda$. Thus it can be expanded in
powers of $p /\Lambda$ with the
higher order terms decreasing in importance: this corresponds to a local
operator
product expansion in the domain of analyticity, equivalent
to a local nonrenormalizable Lagrangian which can be treated as
an honest-to-goodness local field theory. However in general an infinite
series of terms of increasingly higher dimension are generated by matching at
tree level. These cause no problem when the scales are well separated, because
their effects quickly become negligible. But if there are two or more scales
close together (such as $m_t$ and $M_Z$ may be), then we may not be justified
in ignoring terms at higher orders in the expansion. How does one understand
how to interpolate smoothly
between the well-understood situation in which the scales are very different
and the well-understood situation in which the scales are very close together?
How can we keep track of all the infinite number of higher derivative
operators efficiently? Is it possible to deal with the nonlocal effective
Lagrangian directly, without expanding? In the
particular context of a scalar field theory, we will attempt to
answer questions at the one loop level. The approach will be direct. We will
manipulate the nonlocal interactions as if they are expanded in a momentum
expansion, and then show that the resulting $\beta$ functions for the terms in
the momentum expansion can be collected into integro-differential
renormalization group equations for the nonlocal couplings.

The present paper is organized as follows: in section 2 we calculate the
$\beta$ function for a nonlocal four point coupling arising from a one-loop
graph with two internal lines in nonlocal $\phi^4$ theory. We give
this example before the general results of the following sections in order to
point out the important features of our method. In section 3
the method is then applied to obtain the contribution to the renormalization
group equation for a general class of graphs in an arbitrary massive
nonlocal, non-renormalizable parity invariant scalar field theory.

\section{ Basic Example}

\begin{figure}
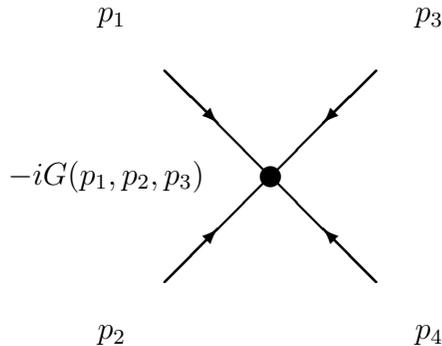

$$\bp(200,160)(-100,-80)
\thicklines
\put(-40,-40){\vector(1,1){20}}
\put(-40,-40){\line(1,1){40}}
\put(-40,40){\vector(1,-1){20}}
\put(-40,40){\line(1,-1){40}}
\put(40,-40){\vector(-1,1){20}}
\put(40,-40){\line(-1,1){40}}
\put(40,40){\vector(-1,-1){20}}
\put(40,40){\line(-1,-1){40}}
\put(60,60){\makebox(0,0){$p_3$}}
\put(-60,60){\makebox(0,0){$p_1$}}
\put(-60,-60){\makebox(0,0){$p_2$}}
\put(60,-60){\makebox(0,0){$p_4$}}
\put(0,0){\circle*{8}}
\put(-25,0){\makebox(0,0)[r]{$-iG(p_1,p_2,p_3)$}}
\ep$$\caption{Basic nonlocal $\phi^4$ vertex.}\end{figure}
The matrix elements in a general nonlocal field theory \cite{NLFTL} are
calculated by writing down a Lagrangian with `smeared' vertices. For instance,
for a nonlocal $\phi^4$ interaction in a scalar
field theory with the discrete symmetry
$\phi \rightarrow -\phi$, the interaction term in the action is
\be
S_4=\int dx_1
dx_2dx_3dx_4F(x_1-x_4,x_2-x_4,x_3-x_4)
\phi(x_1)\phi(x_2)\phi(x_3)\phi(x_4)\label{nl}
\ee
where $x_i$ are spacetime coordinates and $F$ is a nonlocal `form-factor'.
Energy-momentum conservation at each vertex of the corresponding Feynman graph
is expressed in terms of the Fourier transform $G$ of $F$: $G =
G(p_1,p_2,p_3)$ for the basic $\phi^4$ interaction shown in figure 1.
Bose symmetry implies that the nonlocal coupling $G$ is symmetric and
satisfies
\be
G(p_1,p_2,p_3)=G(p_1,p_2,-p_1-p_2-p_3)\,.
\label{g}
\ee
Of course, Lorentz invariance dictates that in the final expression for the
matrix element in momentum space, only scalar products of momenta will appear
as arguments of $G$. {\it Implicit in the definition of $G$ is a mass scale
$\Lambda$ which sets the limit for the region of analyticity of
$G$}, and
for characteristic momenta $p < \Lambda$, $G$ is analytic. We go to the
effective low energy theory by expanding
in a Taylor expansion in $p / \Lambda$. One can think of this Taylor expansion
as the formal implementation of a local operator product expansion of $G$. At
this point, in going to the effective low energy theory, we have actually
changed the high energy behavior of the theory so that integrals which were
convergent in the full theory are divergent in the effective theory. This
trades logs of $\Lambda$ in the full theory for anomalous dimensions in the
low energy theory. This trade allows us to calculate the logs more simply and
to sum them using the the renormalization group.

\begin{figure}
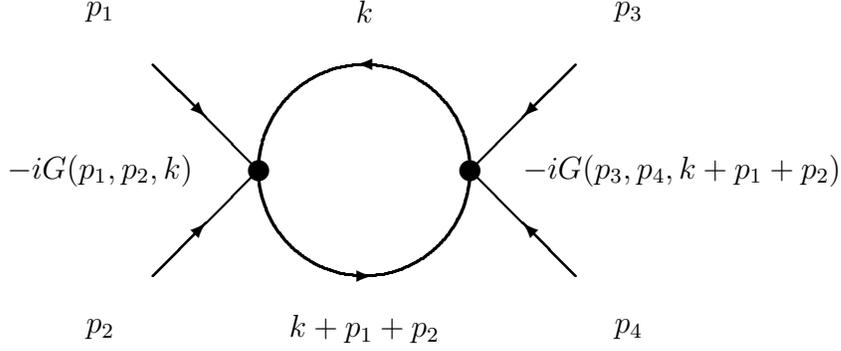

$$\bp(200,160)(-100,-80)
\thicklines
\put(0,0){\ptcommand{\setplotsymbol ({\small .})
\circulararc 360 degrees from 40 0 center at 0 0}}
\put(-80,-40){\vector(1,1){20}}
\put(-80,-40){\line(1,1){40}}
\put(-80,40){\vector(1,-1){20}}
\put(-80,40){\line(1,-1){40}}
\put(80,-40){\vector(-1,1){20}}
\put(80,-40){\line(-1,1){40}}
\put(80,40){\vector(-1,-1){20}}
\put(80,40){\line(-1,-1){40}}
\put(2,40){\vector(-1,0){5}}
\put(-2,-40){\vector(1,0){5}}
\put(100,60){\makebox(0,0){$p_3$}}
\put(-100,60){\makebox(0,0){$p_1$}}
\put(-100,-60){\makebox(0,0){$p_2$}}
\put(100,-60){\makebox(0,0){$p_4$}}
\put(0,-60){\makebox(0,0){$k+p_1+p_2$}}
\put(0,60){\makebox(0,0){$k$}}
\put(-40,0){\circle*{8}}
\put(40,0){\circle*{8}}
\put(-65,0){\makebox(0,0)[r]{$-iG(p_1,p_2,k)$}}
\put(60,0){\makebox(0,0)[l]{$-iG(p_3,p_4,k+p_1+p_2)$}}
\ep$$\caption{Feynman graph contributing to the renormalization of
$G$. }\end{figure}
Let us first discuss as an example the renormalization of the nonlocal
$\phi^4$ interaction from the graph shown in figure 2. For massless fields,
this is the only contribution in one loop. In the massive case, there is also
a contribution from a tadpole graph. In addition, in either case, there are
renormalization of $\phi^{2k}$ interactions for $k>2$. We will systematically
consider them in the next section.

The Feynman integral for the graph in figure 2 is
\be
-\int{d^4k\over(2\pi)^4}\,{G(p_1,p_2,k)\,G(p_3,p_4,k+p_1+p_2)
\over [k^2-m^2+i\epsilon][(k+p_1+p_2)^2-m^2+i\epsilon]}
\label{feynman}
\ee
Combine denominators
\be\ba{c}
{1\over [k^2-m^2+i\epsilon][(k+p_1+p_2)^2-m^2+i\epsilon]}
\cra=\int_0^1d\alpha
{1\over [(1-\alpha)k^2+\alpha(k+p_1+p_2)^2-m^2+i\epsilon]^2}
\ea
\label{combine}
\ee
and shift momenta
\be
k=\ell-\alpha(p_1+p_2)
\label{shift}
\ee
to obtain
\be
-\int_0^1\,d\alpha\,
\int{d^4\ell\over(2\pi)^4}\,{G\bigl(p_1,p_2,\ell
-\alpha(p_1+p_2)\bigr)\,G\bigl(p_3,p_4,\ell+
(1-\alpha)(p_1+p_2)\bigr)
\over \bigl[\ell^2+\alpha(1-\alpha)(p_1+p_2)^2-m^2+i\epsilon\bigr]^2}
\label{feynman2}
\ee
{\bf Now, here is the crucial point. We get the low energy effective theory by
expanding the $G$s in a momentum expansion. An equivalent procedure is to
treat the $G$s as if they were analytic everywhere in momentum space.}
Thus we can deal with the $\ell$ dependence of
$G$ by writing a symbolic Taylor expansion\footnote{Analyticity of the
nonlocal couplings may alternatively be exploited by formally Laplace
transforming in the loop dependent arguments, which yields instead of
$e^{\ell \cdot {\partial \over \partial q}} G$ an exponential $e^{-\ell \cdot
s} \tilde{G}$ in terms of the Laplace transform of $G$, along with an extra
Laplace inversion integral. The rest of the computation is essentially
identical.} {\bf Here we are effectively doing the momentum expansion. But the
key is that we can resum the final result into an finite integral over the
original nonlocal couplings.}
 \be\ba{c} G\bigl(p_1,p_2,\ell-
\alpha(p_1+p_2)\bigr)\,G\bigl(p_3,p_4,\ell+
(1-\alpha)(p_1+p_2)\bigr)\cra
=e^{\ell{\partial\over\partial q}}\left[G\bigl(p_1,p_2,q\bigr)\,
G\bigl(p_3,p_4,q+p_1+p_2
\bigr)\right]_{q=-\alpha(p_1+p_2)}\ea
\label{symbolic}
\ee

Now all the dependence on the loop momentum is in the denominators and in the
exponential, so we just have to do the Feynman integral of the exponential
\be
\int{d^4\ell\over(2\pi)^4}\,{e^{\ell{\partial\over\partial q}}
\over\bigl[\ell^2+\alpha(1-\alpha)(p_1+p_2)^2-m^2+i\epsilon\bigr]^2}
\label{exponential}
\ee
or to be more precise, the dimensionally regularized integral
\be
\int{d^{4-\epsilon}\ell
\over(2\pi)^{4-\epsilon}\mu^{-2\epsilon}}\,
{e^{\ell{\partial\over\partial q}}
\over\bigl[\ell^2+\alpha(1-\alpha)(p_1+p_2)^2-m^2+i\epsilon\bigr]^2}
\label{dr}
\ee

We will do this by manipulating the exponential like a power series, because
the basic assumption is analyticity in the momenta. It is at this point that
we have irrevocably changed the high energy behavior and gone over the
effective low energy theory. Because of symmetry, only
even terms in $\ell$ contribute,

\be
\sum_{r=0}^\infty\,{1\over (2r)!}\,\int{d^{4-\epsilon}\ell
\over(2\pi)^{4-\epsilon}\mu^{-2\epsilon}}\,
{\left(\ell{\partial\over\partial q}\right)^{2r}
\over\bigl[\ell^2+\alpha(1-\alpha)(p_1+p_2)^2-m^2+i\epsilon\bigr]^2}
\label{expand}
\ee
We calculate this as follows
\be \ba{c}\ds \int\,{d\Omega_\ell\over\Omega}\,
\ell_{\mu_1}\ell_{\mu_2}\cdots\ell_{\mu_{2r}}\\ \ds
=A_r\,\left(\ell^2\right)^{r}\,\overbrace{\left[
g_{\mu_1\mu_2}\cdots g_{\mu_{2r-1}\mu_{2r}}
+ \mbox{perms}\right]}^{(2r)!/(2^rr!)\;\rm terms}\,.\ea\ee
where
\be
\Omega \equiv \int d\Omega_\ell\,.
\ee
Contracting with $g^{\mu_1\mu_2}$ gives
\be A_{r-1}=\left(4+2(r-1)\right)\,A_r
=2(r+1)\,A_r\ee
or
\be A_r={1\over 2^r (r+1)!}\ee
and thus
\be \ba{c}{1\over (2r)!}\int\,{d\Omega_\ell\over\Omega}\,
\left(\ell{\partial\over\partial q}\right)^{2r}
={1\over 4^rr!(r+1)!}\,\left[\ell^2\right]^r
\,\left[\left({\partial\over\partial q}\right)^2\right]^r\,.
\label{average}
\ea\ee
So the integral is
\be
\sum_{r=0}^\infty\,{1\over 4^rr!(r+1)!}\,\int{d^{4-\epsilon}\ell
\over(2\pi)^{4-\epsilon}\mu^{-2\epsilon}}\,
{\left[\ell^2\right]^r
\,\left[\left({\partial\over\partial q}\right)^2\right]^r
\over\bigl[\ell^2+\alpha(1-\alpha)(p_1+p_2)^2-m^2+i\epsilon\bigr]^2}
\label{integral}
\ee
which we can calculate in the usual way. Wick rotate
\be
i\sum_{r=0}^\infty\,{(-1)^r\over 4^rr!(r+1)!}\,\int{d^{4-\epsilon}\ell
\over(2\pi)^{4-\epsilon}\mu^{-2\epsilon}}\,
{\left[\ell^2\right]^r
\,\left[\left({\partial\over\partial q}\right)^2\right]^r
\over\bigl[\ell^2+A(\alpha)^2\bigr]^2}
\label{integral1}
\ee
where
\be
A(\alpha)^2=-\alpha(1-\alpha)(p_1+p_2)^2+m^2\, ,
\label{bha}
\ee
and $A(\alpha)^2$ is positive for Euclidean momenta.
Now doing the $\ell$ integral gives
\be
{2i\over(4\pi)^2\epsilon\mu^{-2\epsilon}}\,
\,\sum_{r=0}^\infty\,{1\over4^rr!^2}
\,A(\alpha)^{2r}
\,\left[\left({\partial\over\partial q}\right)^2\right]^r+\cdots
\label{integral2}
\ee
where we have dropped everything but the $1/\epsilon$ pole and the associated
$\mu$ dependence.
We can write the result as
\be
{2i\over(4\pi)^2\epsilon\mu^{-2\epsilon}}\, {\partial\over\partial
x}\, \left[ \sum_{r=0}^\infty\,{x\over4^rr!(r+1)!}
\,[-xA(\alpha)^2]^{r}
\,\left[\left({\partial\over\partial q}\right)^2\right]^r \right]_{x=1}.
\label{integral3}
\ee
This can then be turned back into the (Euclidean) angular average of an
exponential, just inverting \eqr{average},
\be
{2i\over(4\pi)^2\epsilon\mu^{-2\epsilon}}\,{\partial\over\partial
x}\,\left[\int{d\Omega_e\over\Omega}\,
\,x \, e^{\sqrt{x}A(\alpha)\,
e{\partial\over\partial q}}\right]_{x=1}
\label{integral4}
\ee
where $e$ is a Euclidean unit vector.
Now that we have done the $\ell$ integral and extracted the $1/\epsilon$ pole,
we can use the Taylor series to put the form \eqr{integral4} in terms of the
 $G$s. The result is
\be\ba{c}
{i\over(8\pi)^2\epsilon\mu^{-2\epsilon}}\,{\partial\over\partial
x}\,\Biggl[\int{d\Omega_e\over\Omega}\,\int_0^1d\alpha \, x
\left[G\bigl(p_1,p_2,\sqrt{x}A(\alpha)e-
\alpha(p_1+p_2)\bigr)\right.\cra
\left. G\bigl(p_3,-p_1-p_2-p_3,
\sqrt{x}A(\alpha)e+
(1-\alpha)(p_1+p_2)\bigr)\right] \Biggr]_{x=1}\ea
\label{integral5}
\ee
with the derivative evaluated at $x=1$.
Thus we have reduced the divergent part of the original Feynman graph to an
integral of the $G$s and their derivatives over {\bf finite ranges} (for
Euclidean momenta). These last integrals are well-behaved and
 could be done numerically. Hence, with $\Omega = 2 \pi^2$ in for Euclidean
four space, we obtain
 the one-loop $\beta$ function for the nonlocal interaction, $G$:
\be\ba{c}
\beta_{G(p_1,p_2,p_3)}=
{1\over 16\pi^4}\,{\partial\over\partial
x}\, \Biggl[ \int{d\Omega_e}\,\int_0^1d\alpha \, x
\left[G\bigl(p_1,p_2,\sqrt{x}A(\alpha)e-\alpha(p_1+p_2)\bigr)\right.\cra
\left. G\bigl(p_3,-p_1-p_2-p_3,\sqrt{x}A(\alpha)e+
(1-\alpha)(p_1+p_2)\bigr)\right] \Biggr]_{x=1}+{\rm cross\atop terms}\ea
\label{beta}
\ee
Now in the massless case, where there are no other contributions to the
renormalization of the $\phi^4$ coupling, the ``running'' nonlocal coupling
satisfies the integrodifferential renormalization group equation
\be
\mu{\partial\over\partial \mu}\,G(p_1,p_2,p_3)
=\beta_{G(p_1,p_2,p_3)}.
\label{rge}
\ee
A useful check on this result is obtained by going to the `local limit', i.e.
$G \rightarrow g$ where $g$ is the usual coupling constant of local $\phi^4$
theory. Indeed, we know that the contribution to the $\beta$ function in
local $\phi^4$ theory at one loop is
\be
\beta(g) = {3 g^2 \over 16 \pi^2} >0.
\ee
Substituting $G = g$, inserting
an explicit symmetry factor of $1/2$, and summing over the crossed graphs,
\eqr{beta} indeed yields
\be
\beta_{G(p_1,p_2,p_3)} \rightarrow \beta(g)
\quad {\rm as} \quad G \rightarrow g.
\ee

Now the nonlocal quartic coupling $G$ can induce changes in the coupling
terms with more fields. For instance, the one-loop diagram relevant to the
renormalization of the $\phi^6$ coupling is shown in figure 3.
\begin{figure}
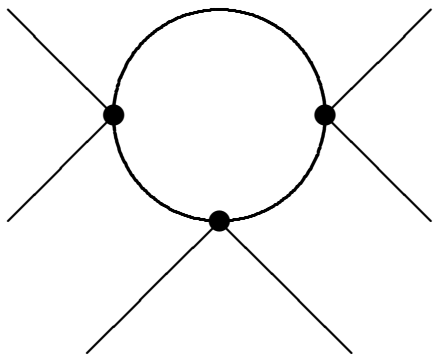

$$\bp(200,160)(-100,-100) \thicklines
\put(0,0){\ptcommand{\setplotsymbol ({\small .})
\circulararc 360 degrees from 40 0 center at 0 0}}
\put(-80,-40){\line(1,1){40}}
\put(-80,40){\line(1,-1){40}}
\put(80,-40){\line(-1,1){40}}
\put(80,40){\line(-1,-1){40}}
\put(-40,0){\circle*{8}}
\put(40,0){\circle*{8}}
\put(0,-40){\circle*{8}}
\put(0,-40){\line(1,-1){50}}
\put(0,-40){\line(-1,-1){50}}
\ep$$\caption{Feynman graph contributing to the running of
$\phi^6$ coupling.}\end{figure}

For massless fields, this does not mix back
into $G$ in a mass independent renormalization scheme, because the relevant
Feynman graph of figure 4 vanishes.
However, in the massive case, the graph of figure 4 gives a non-vanishing
contribution. We will now systematically compute the $\beta$ functions of
the general nonlocal couplings in a massive nonrenormalizable theory.
\begin{figure}
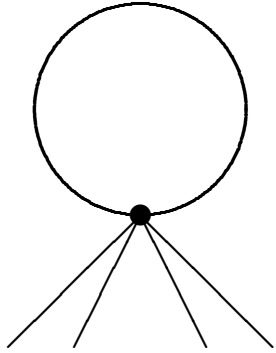

$$\bp(200,160)(-100,-100)
\thicklines
\put(0,0){\ptcommand{\setplotsymbol ({\small .})
\circulararc 360 degrees from 40 0 center at 0 0}}
\put(0,-40){\circle*{8}}
\put(0,-40){\line(1,-1){50}}
\put(0,-40){\line(1,-2){25}}
\put(0,-40){\line(-1,-1){50}}
\put(0,-40){\line(-1,-2){25}}
\ep$$\caption{Feynman graph that vanishes for massless $\phi$s.}\end{figure}

\section{ General Results}

Working within the formalism of a massive nonlocal effective
theory induced by some unknown full theory, we are forced to consider an
effective Lagrangian with operators with an arbitrary number of low energy
fields. In this section, we will compute the
one-loop running of a general non-renormalizable,
nonlocal scalar effective theory with $\phi\rightarrow-\phi$ symmetry.
Specifically, our purpose is to
isolate the dimensional regularization pole of a $2m$ point function of type
$n$ (i.e. with $n$ vertices or propagators) at one loop.
The plan is as follows: we first construct the expressions corresponding to
the Feynman integral of a general graph with a special choice of momentum
labelling conventions.
We then describe the application of the method of the last section to two
relevant examples: the tadpole renormalization of the nonlocal `$\phi^4$'
coupling with a $\phi^6$ operator insertion, and the renormalization of a
nonlocal
$\phi^6$ coupling with three $\phi^4$ operator
 insertions. The last example is
useful in highlighting the special features of renormalizing graphs with
more than two internal lines.
  Finally, we attack the general case, and give the complete
expression for the $1/ \epsilon$ pole of the $2m$ point function of type $n$
at one loop as a surprisingly compact
multi-Feynman-parameter, multi-dimensional angular integral.

\subsection{\normalsize \bf Preliminaries}
We assume that the dimensionally continued $d=4-\epsilon$ dimensional
non-local Lagrangian has a $\phi\rightarrow-\phi$ symmetry and hence has an
interaction term proportional to
\be \cL_{\rm int.} = \sum_{r=1}^{\infty} {\mu}^{\epsilon(r-1)} G_{2r}
\phi^{2r}. \label{coupling} \ee
The mass scale $\mu$ is introduced to keep the dimensions of the nonlocal
couplings $G_{2r}$ fixed under dimensional continuation. The interaction is
not normal ordered\footnote{Hence fields at the same point can be contracted,
and tadpoles will occur explicitly.}, and each $G_{2r}$ is some nonlocal
function which is analytic in the region under consideration, depends as a
consequence of momentum conservation on $2r-1$ linearly independent momenta,
and may have dimensions proportional to some power of an implicit scale of
nonlocality $\Lambda$.

\begin{figure}
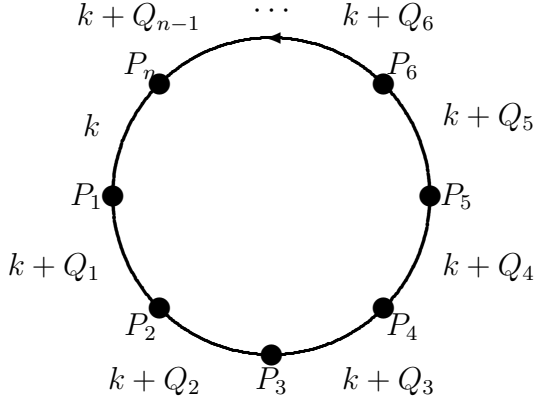

$$\bp(200,160)(-100,-80)
\thicklines
\put(0,0){\ptcommand{\setplotsymbol ({\small .})
\circulararc 360 degrees from 60 0 center at 0 0}}
\put(0,-60){\circle*{8}}
\put(-60,0){\circle*{8}}
\put(60,0){\circle*{8}}
\put(42.4,42.4){\circle*{8}}
\put(-42.4,42.4){\circle*{8}}
\put(42.4,-42.4){\circle*{8}}
\put(-42.4,-42.4){\circle*{8}}
\put(-64.7,26.8){\makebox(0,0)[r]{$k$}}
\put(-64.7,-26.8){\makebox(0,0)[r]{$k+Q_1$}}
\put(-26.8,-64.7){\makebox(0,0)[rt]{$k+Q_2$}}
\put(26.8,-64.7){\makebox(0,0)[lt]{$k+Q_3$}}
\put(-26.8,64.7){\makebox(0,0)[rb]{$k+Q_{n-1}$}}
\put(64.7,-26.8){\makebox(0,0)[l]{$k+Q_4$}}
\put(64.7,26.8){\makebox(0,0)[lb]{$k+Q_5$}}
\put(26.8,64.7){\makebox(0,0)[lb]{$k+Q_{6}$}}
\put(-49.5,49.5){\makebox(0,0){$P_n$}}
\put(-49.5,-49.5){\makebox(0,0){$P_2$}}
\put(49.5,-49.5){\makebox(0,0){$P_4$}}
\put(49.5,49.5){\makebox(0,0){$P_6$}}
\put(-70,0){\makebox(0,0){$P_1$}}
\put(0,-70){\makebox(0,0){$P_3$}}
\put(70,0){\makebox(0,0){$P_5$}}
\put(0,70){\makebox(0,0){$\cdots$}}
\put(2.5,60){\vector(-1,0){5}}
\ep$$\caption{Type-$n$ Feynman graph. The blobs signify arbitrary $\phi^{2r}$
insertions.}\end{figure}

A diagram with $n$ internal lines will be called type-$n$. At one loop, a
type-$n$ graph has $n$ vertices connected to external lines. Let
\be
N=\sum_{i=1}^n v_i\,,
\ee
where $v_i$ is the number of external lines emanating from the $i$th vertex.
The loop integral for the type-$n$ one-loop renormalization of the $N$ point
function (shown in figure 5) is
\be
{I= {\int}{d^4 k \over (2 \pi)^4}} {{\prod_{i=1}^n
G^i_{v_i+2} (
\ldots, k +Q_{i-1})}
\over
{ (k^2) (k+Q_1)^2 \ldots (k+Q_{n-1})^2}}.
\label{integrand}
\ee
where the upper index on the $G's$ is used to number the vertices as one goes
along the loop.

Here all external momenta are taken to be incoming (signified by
$\ldots$, in the numerator, different for different $G^i$) and we have
defined \bea Q_0 & = & 0 \label{momdef} \\ \nonumber
Q_1 & = & P_1 \\ \nonumber
Q_2 & = & P_2 + Q_1 \\ \nonumber
Q_3 & = & P_3 + Q_2 \\ \nonumber
\vdots & \vdots & \vdots \\ \nonumber
Q_{i} & = & P_i + Q_{i-1} = \sum_{j=1}^{j=i} P_j \\ \nonumber
Q_{n} & = & Q_{n-1} + P_n = 0,
\eea
and $P_i$ is the sum of the external momenta flowing into the
$i$th vertex.
Energy momentum conservation is simply $Q_n = 0$. Symmetry factors are
suppressed.

Using this notation, the loop integral for the
for the $\phi^4$
coupling that we computed in the last section is obtained by setting $n=2, v_1
=2, v_2=2$:
\be
{I= {\int}{d^4 k \over (2 \pi)^4}} {G^1_4(p_1,p_2,k)
G^2_4(p_3,p_4,k+Q_1)
\over (k^2 + i \epsilon) \left[ (k + Q_1)^2 + i \epsilon
\right] }.
\ee

Using the generalized Feynman identity
\bea
& & {1 \over A_1^{\rho_1} A_2^{\rho_2} \ldots A_n^{\rho_n} } =
{\Gamma(\rho_1 +
\rho_2 + \ldots + \rho_n) \over \Gamma(\rho_1) \Gamma(\rho_2) \ldots
\Gamma(\rho_n)} \label{feynmanpar} \\
& & \times \int_0^1 d \alpha_1 \ldots d \alpha_n
{\alpha_1^{\rho_1 -1} \alpha_2^{\rho_2 -1} \ldots \alpha_n^{\rho_n-1} \delta(1
-\alpha_1-\alpha_2-\ldots -\alpha_n) \over (\alpha_1 A_1 + \alpha_2 A_2 +
\ldots + \alpha_n A_n)^{\rho_1 + \rho_2 + \ldots + \rho_n}}, \nonumber
\eea
 combine the
denominators
\bea
& & { 1 \over {(k^2) (k+Q_1)^2 \ldots (k+Q_{n-1})^2}} \label{combine2} \\
\nonumber
&= & (n-1)! \int \prod_i
d \alpha_i
{1 \over { {[} k^2 (1 - \sum_{i=1}^{n-1} \alpha_i) +
\sum_{r=1}^{n-1} (k
+Q_r)^2 \alpha_r {]}^n} } ,
\eea
and shift the loop momentum
\be
k = \ell - \sum_{s=1}^{n-1} Q_s \alpha_s
\label{shift2}
\ee
so that
\be
k^2 = \ell^2 + (\sum_{s=1}^{n-1} Q_s \alpha_s)^2 - 2 \ell\cdot
\sum_{s=1}^{n-1} Q_s \alpha_s,
\ee
which, when put into the denominator gives
\bea
{\rm D} & = & {[} \ell^2 + (\sum_{s=1}^{n-1} Q_s \alpha_s)^2 -2 \ell
\cdot
\sum_{s=1}^{n-1} Q_s \alpha_s
-\ell^2 \sum_{t=1}^{n-1} \alpha_t \label{denominator} \\ \nonumber
& & - (\sum_{s=1}^{n-1} Q_s \alpha_s)^2 \sum_{t=1}^{n-1} \alpha_t
+( 2 \ell \cdot \sum_{s=1}^{n-1} Q_s \alpha_s)(\sum_{t=1}^{n-1}
\alpha_t)
+ \sum_{r=1}^{n-1} \ell^2 \alpha_r \\ \nonumber
& & + \sum_{r=1}^{n-1}(\sum_{s=1}^{n-1} Q_s \alpha_s)^2 \alpha_r
-2 \sum_{r=1}^{n-1} \ell \cdot \sum_{s=1}^{n-1} Q_s \alpha_s
\alpha_r
+ \sum_{r=1}^{n-1} Q_r^2 \alpha_r \\ \nonumber
& & + 2 \sum_{r=1}^{n-1} \ell \cdot Q_r \alpha_r
- 2 \sum_{r=1}^{n-1} Q_r (\sum_{s=1}^{n-1} Q_s \alpha_s) \alpha_r
{]}^n .
\eea
After cancellation \eqr{denominator} yields the denominator
\be
{\rm D} = {[} \ell^2 - A^2 {]}^n
\ee
where
\be
A^2 \equiv (\sum_{s=1}^{n-1} Q_s \alpha_s)^2 - \sum_{r=1}^{n-1} Q_r^2
\alpha_r.
\label{asquare}
\ee
For the \em massive \rm case, we simply substitute in \eqr{combine2}
\bea
k^2 & \rightarrow & k^2 -m^2 \\
(k+Q_r)^2 & \rightarrow & (k+Q_r)^2 - m^2
\eea
so that we just get an additional term in \eqr{denominator}
 when combining denominators after the shift \eqr{shift2}:
\be
-m^2(1-\sum_{s=1}^{n-1}\alpha_s) + \sum_{s=1}^{n-1} (-m^2)
\alpha_r = -m^2.
\ee
So the general form for the denominator of the right hand side
of \eqr{combine2} with massive fields is
\be
D = {[} k^2 - (\sum_{s=1}^{n-1} Q_s \alpha_s)^2 +
(\sum_{r=1}^{n-1} Q_r^2
\alpha_r) - m^2 {]}^n.
\ee
Now the shift \eqr{shift2} changes the argument of the numerator factors in
\eqr{integrand} also,
to yield the final expression for the Feynman integral\footnote{
The reader may check that with $n=2, v_1=v_2=2$, the above relations give the
correct integral for the example of the first section:
\be
{\int d \alpha} \int {d^d\ell\over (2 \pi)^d}
{G_4(p_1,p_2,\ell-\alpha(p_1+p_2)) G_4(p_3,p_4,\ell+(1-\alpha)(p_1+p_2)) \over
\left[ \ell^2 + \alpha(1 - \alpha) Q_1^2 -m^2 \right]^2 }
\ee
with $Q_1 = p_1 + p_2$.}
\be
(n-1)! \int \prod_{j=1}^{n-1} d \alpha_{j}
\int {d^{d}\ell \over (2 \pi)^d}
{{{\prod_{i=1}^n G^i_{v_i+2} ( \ldots, \ell - \sum_{s=1}^{n-1} Q_s
\alpha_s
+Q_{i-1})}}
\over
{{[} \ell^2 - (\sum_{s=1}^{n-1} Q_s \alpha_s)^2 + (\sum_{r=1}^{n-1}
Q_r^2
\alpha_r) - m^2 + i \epsilon {]}^n }}
\ee

\em Important remarks on notation \rm :
(1) In intermediate steps of the computations, we denote the dependence of
each nonlocal function on the (distinct!) external momenta by ellipsis. This
is useful since the external momenta play a trivial part in the loop
integration, and may be reinstated by examination in the final expressions.
(2) Also, since at one loop any vertex shares two and only two lines with the
loop, by energy momentum conservation the loop momentum appears only once as
an argument of any $G^i$ and will be put in its last slot. (3) The first
vertex will have only the loop momentum as its last entry.

\em Crossing \rm :
The external momenta can be exchanged amongst themselves. The final result,
however, depends only on the number of distinct Lorentz invariants of the four
momenta $p_i$ ($i=2m$ for renormalization of the $2m$ point function) under
the condition $\sum_{i=1}^{2m} p_i = 0$ and $p_i^2 = m^2$. This equals the
total number of graphs related by `crossing' where the crossed graphs can be
obtained by exchanging external momenta. We will give explicit expressions
only for one member of each crossed set.

\em Counting $i'$s \rm : Ignoring for the moment the $i'$s appearing due to
Wick rotation and integration (see below), the integrand itself gives no
powers of $i$. This is seen as follows: each propagator is $i \over
p^2-m^2+i\epsilon$, and each nonlocal vertex has a Feynman rule $-i G_{2r}$.
Since the number of vertices equals the number of internal lines at one loop,
for a type-$n$ graph we obtain $(-i)^n i^n = (-1)^n (i)^{2n} = (-1)^{2n} =
1$. So the only source of $i$'s and minus signs is
the integration formula.

\em Computational Tools \rm:

In order to do the dimensionally regularized loop integrals we will need the
well-known integration formula (for Euclidean momenta):
\be
\int d^dq \, {(q^2)^r \over (q^2-A^2)^{n}}
= i\pi^{d/2} (-1)^{n+r} (A^2)^{r-n+{d \over 2}}
{\Gamma( r+ {d \over 2}) \Gamma(n-r-{d \over 2}) \over \Gamma({d
\over 2})
\Gamma(n)},
\label{key}
\ee
and the expansion
\be
\Gamma(-n+\epsilon) = { (-1)^n \over n!} [ {1 \over \epsilon} + (1 + {1\over
2} + \ldots + {1 \over n} - \gamma) + \cO (\epsilon) ] ,
\label{gam}
\ee
where $\gamma = 0.5772157$ is Euler's constant.

Also note that in $2 \kappa -2$ dimensions ($\kappa$ a positive integer)
\be \ba{c}\ds\int\,{d\Omega^{2
\kappa-2}_\ell\over\Omega^{2 \kappa-2}}\,
\ell_{\mu_1}\ell_{\mu_2}\cdots\ell_{\mu_{2r}}\\ \ds
=A^{2\kappa-2}_{r}\,\left(\ell^2\right)^{r}\,\overbrace{\left[
g_{\mu_1\mu_2}\cdots g_{\mu_{2r-1}\mu_{2r}}
+ \mbox{perms}\right]}^{(2r)!/(2^rr!)\;\rm terms,
}
\ea\ee
where we have defined
\be
\Omega^d \equiv \int d\Omega^d={2 \pi^{d/2} \over \Gamma(d/2)}\,.
\label{kangular}
\ee
Contracting with $g^{ \mu_1\mu_2}$ gives
\be A^{2 \kappa-2}_{r-1}=\left[2 \kappa -2 + 2(r-1)\right] \,A^{2 \kappa-2}_r
=(2 \kappa + 2r -4) \,A^{2 \kappa-2}_{r}
\ee
So
\be A^{2 \kappa-2}_r={1\over 2^r (\kappa+r-2)!}
\ee
and thus in $2 \kappa-2$ dimensions
\be \ba{c}{1\over (2r)!}{d\Omega^{2 \kappa-2}_\ell\over\Omega^{2
\kappa-2}}\, \left(\ell{\partial\over\partial
q}\right)^{2r} \cra={1\over 4^r r!(r+\kappa-2)!}\,\left[\ell^2\right]^r
\,\left[\left({\partial\over\partial q}\right)^2\right]^r.
\label{averaged}
\ea\ee
The importance of the result of \eqr{averaged} will be seen when we isolate
the pole pieces appearing from graphs with more than two internal lines. The
point is that the leading terms in the Taylor expansion then give Feynman
integrals which are manifestly convergent by power counting, and divergences
appear only at some higher order in the Taylor expansion. This implies that
the sum of dimensional regularization poles does not begin at zero, and we
cannot immediately write the result as an angular integral over a Euclidean
unit vector in four dimensions, as we did in the last section. One choice is
to redefine the infinite sum to start at zero and compensate for the
redefinition by subtracting off a finite sum (which vanishes under the action
of differential operators in the dummy parameters). Alternatively, using
\eqr{averaged}, we can re-sum the poles for a graph with $n>2$ internal lines,
\bf without reference to any dummy variables \rm, in terms of a $2n-2$
dimensional angular integral! The example of the $2-2-2$ graph below will give
the explicit details of how this is done.

The $\phi^6$ operator can be renormalized at one loop by a type-$1$ (tadpole)
with a $G_8$ coupling insertion, a type-$2$ graph with $G_6, G_4$ insertions,
or the most convergent graph, of type-$3$, with three $G_4$ insertions. For
the last one, the relevant
graph with incoming external momenta is given in figure 6.
\begin{figure}
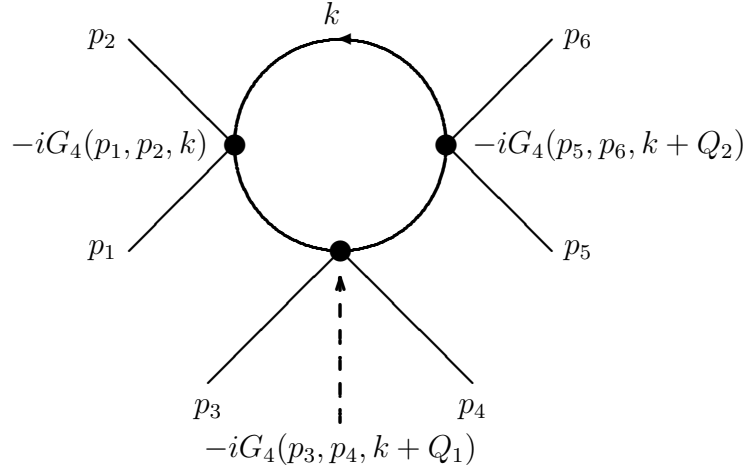

$$\bp(200,170)(-100,-110)
\thicklines
\put(0,0){\ptcommand{\setplotsymbol ({\small .})
\circulararc 360 degrees from 40 0 center at 0 0
\setdashes\arrow <5pt> [.3,.6] from 0 -105 to 0 -50}}
\put(-80,-40){\line(1,1){40}}
\put(-80,40){\line(1,-1){40}}
\put(80,-40){\line(-1,1){40}}
\put(80,40){\line(-1,-1){40}}
\put(-40,0){\circle*{8}}
\put(40,0){\circle*{8}}
\put(0,-40){\circle*{8}}
\put(0,-40){\line(1,-1){50}}
\put(0,-40){\line(-1,-1){50}}
\put(-90,-40){\makebox(0,0){$p_1$}}
\put(-90,40){\makebox(0,0){$p_2$}}
\put(-50,-100){\makebox(0,0){$p_3$}}
\put(50,-100){\makebox(0,0){$p_4$}}
\put(90,-40){\makebox(0,0){$p_5$}}
\put(90,40){\makebox(0,0){$p_6$}}
\put(-50, 0){\makebox(0,0)[r]{$-iG_4(p_1,p_2,k)$}}
\put(50, 0){\makebox(0,0)[l]{$-iG_4(p_5,p_6,k+Q_2)$}}
\put(0, -115){\makebox(0,0){$-iG_4(p_3,p_4,k+Q_1)$}}
\put(0,50){\makebox(0,0)[r]{$k$}}
\put(3,40){\vector(-1,0){5}}
\ep$$\caption{Type-$3$ Feynman graph contributing to
$\beta_{G_6}$.}\end{figure}

\begin{figure}
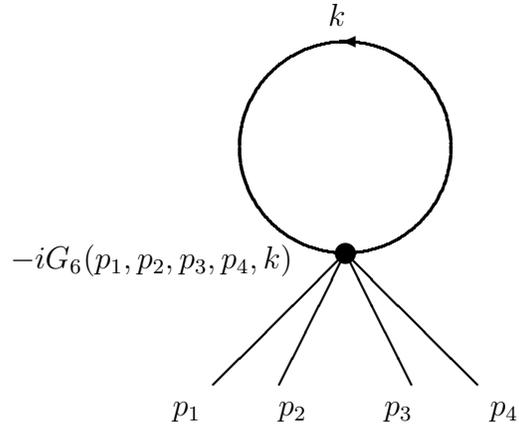

$$\bp(200,160)(-100,-100)
\thicklines
\put(0,0){\ptcommand{\setplotsymbol ({\small .})
\circulararc 360 degrees from 40 0 center at 0 0}}
\put(0,-40){\circle*{8}}
\put(0,-40){\line(1,-1){50}}
\put(0,-40){\line(1,-2){25}}
\put(0,-40){\line(-1,-1){50}}
\put(0,-40){\line(-1,-2){25}}
\put(-60,-100){\makebox(0,0){$p_1$}}
\put(-20,-100){\makebox(0,0){$p_2$}}
\put(20,-100){\makebox(0,0){$p_3$}}
\put(60,-100){\makebox(0,0){$p_4$}}
\put(-20,-43){\makebox(0,0)[r]{$-iG_6(p_1,p_2,p_3,p_4,k)$}}
\put(0,50){\makebox(0,0)[r]{$k$}}
\put(3,40){\vector(-1,0){5}}
\ep$$\caption{Tadpole graph that vanishes for
massless $\phi$s.}\end{figure}

As mentioned in the last section, for massless fields this will not mix back
into $G_4$ in a mass independent renormalization scheme, because
the Feynman graph in figure 7 vanishes.
However, for massive light fields the tadpole graph does not vanish, and we
will therefore evaluate this type-$1$ `tadpole' graph first.

\subsection{\normalsize \bf Tadpole diagram contribution to $\beta_{G_4}$}
We need to do the integral (see figure 7), where $r=1$ in \eqr{coupling}, and
$n=1, v_1=4, p_4=-(p_1+p_2+p_3)$,
\be
 \mu^{ 2 \epsilon} \int {d^d k \over (2 \pi)^d} {G_6(p_1,p_2,p_3,p_4, k)
\over (k^2 -m^2+ i \epsilon)}.
\ee
No momentum shift is needed since the denominator is purely quadratic in the
loop momentum. Now because $G_6$ is analytic, we can deal with the $k$
dependence by writing a symbolic Taylor expansion
\be
 G_6 \bigl(p_1,p_2,p_3,p_4,k \bigr)
=e^{k{\partial\over\partial q}}
\left[G_6\bigl(p_1,p_2,p_3,p_4,q\bigr)\right]_{q=0}
\label{symbolic2}
\ee

Then we just have to do the dimensionally regularized integral
\be
\int{d^{4-\epsilon}k
\over(2\pi)^{4-\epsilon}\mu^{-2\epsilon}}\,
{e^{k{\partial\over\partial q}}
\over\bigl[k^2-m^2+i\epsilon\bigr]}.
\label{dr2}
\ee

We again do this by manipulating the exponential like a power series, because
the basic assumption is analyticity in the momenta. Because of symmetry, only
even terms in $k$ contribute, so the integral to be done equals

\be
\sum_{r=0}^\infty\,{1\over (2r)!}\,\int{d^{4-\epsilon}k
\over(2\pi)^{4-\epsilon}\mu^{-2\epsilon}}\,
{\left(k{\partial\over\partial q}\right)^{2r}
\over\bigl[k^2-m^2+i\epsilon\bigr]}
\label{expand2}
\ee

which finally yields the integral (just as in example of last section)
\be
\sum_{r=0}^\infty\,{1\over 4^rr!(r+1)!}\,\int{d^{4-\epsilon}k
\over(2\pi)^{4-\epsilon}\mu^{-2\epsilon}}\,
{\left[k^2\right]^r
\,\left[\left({\partial\over\partial q}\right)^2\right]^r
\over\bigl[k^2-m^2+i\epsilon\bigr]}.
\label{integraltt}
\ee
After performing a Wick rotation, and doing the integral using \eqr{key}
and \eqr{gam} we obtain for the $1 / \epsilon$ pole and the associated $\mu$
dependence:
\be
{im^2\over(4\pi)^2\epsilon\mu^{-2\epsilon}}\,
\,\sum_{r=0}^\infty\,{1\over 4^r r!(r+1)!}
\,[m^2]^r
\,\left[\left({\partial\over\partial q}\right)^2\right]^r+\cdots
\label{tadpolesum}
\ee

This can then be turned back into the (Euclidean) angular average of an
exponential, by just inverting \eqr{average},
\be
{im^2\over(4\pi)^2\epsilon\mu^{-2\epsilon}}\int\,{d\Omega^4_e\over\Omega^4}
\, e^{m\,e{\partial\over\partial q}}
\label{integralt3}
\ee
where $e$ is a Euclidean unit vector, and $\Omega^4 = 2 \pi^2$.
Now that we have done the loop integral and extracted the $1/\epsilon$ pole,
we can use the Taylor series to put the form \eqr{integralt3} in terms of
$G_6$. The result for the pole part then equals
 \be
{im^2\over 32 \pi^4 \epsilon\mu^{-2\epsilon}} \int\,{d\Omega^4_e}
\left[G_6\bigl(p_1,p_2,p_3,-p_1-p_2-p_3,me)\bigr)\right].
\label{integralt4}
\ee
Thus we have reduced the divergent part of the original Feynman graph
to an integral of $G_6$ over a finite region (for
Euclidean momenta).
Hence we have obtained the tadpole contribution to the one-loop $\beta$
function for the nonlocal interaction, $G_4$:

\bea
\lefteqn { \beta_{G_{4[4]} (p_1,p_2,p_3)} \equiv \mu {d \over d \mu}
G_{4[4]}(p_1,p_2,p_3) } \label{tadpolebeta} \\ \nonumber
 & & =  { m^2 \over 32 \pi^4 }
\int
{d\Omega^4_e} G_6(p_1,p_2,p_3,-p_1-p_2-p_3, m e),
\eea
where the subscript on $G$ on the LHS denotes the coupling that is
renormalized, with the number of external lines at each vertex for the
graph under consideration in square brackets (i.e. type of graph
considered).\footnote{Also note than in the local limit, replacing $G_6$ by a
`constant'
$g_6$, we get
$\beta_{g_{4[4]}} = {m^2 g_6 \over 16 \pi^2}$ which may be directly obtained
in the local limit as a check.}

\subsection{\normalsize \bf 2-2-2 diagram contribution to $\beta_{G_6}$}

The Feynman integral from figure 6, with $Q_1 = p_1+p_2, Q_2 =
p_1+p_2+p_3+p_4$, $\sum_{i=1}^{i=6}p_i=0$; and $v_1=v_2=v_3=2$, equals
\be
{= \mu^{3 \epsilon} \int {d^4 k \over (2 \pi)^4}}
{
G_4(p_1,p_2,k) \, G_4(p_3,p_4, k+Q_1) \, G_4(p_5,p_6, k+Q_2) \over
(k^2-m^2 +i \epsilon) [(k+Q_1)^2 - m^2 +i \epsilon] [(k+Q_2)^2 -
m^2 +i \epsilon]
}.
\ee

Combining denominators and making the shift $k = \ell - (Q_1
\alpha_1 + Q_2
\alpha_2)$, we get
the denominator
\be
D = \left[ \ell^2 - Q_1^2 \alpha_1(\alpha_1-1) - Q_2^2 \alpha_2
(\alpha_2-1) - 2 Q_1
\cdot Q_2 \alpha_1 \alpha_2 - m^2 \right] ^3
\ee
so that the integral becomes (with a factor of $2$ from the Feynman trick)
\bea
& & = 2 \int \prod_i d \alpha_i \int {d^d \ell \over (2 \pi)^d} {1
\over D}
\Bigl[ G_4(p_1,p_2, \ell- (Q_1 \alpha_1 + Q_2 \alpha_2)) \\ \nonumber
& & G_4(p_3,p_4, \ell-
(Q_1 (\alpha_1 - 1 )+ Q_2 \alpha_2)) G_4(p_5,p_6, \ell -Q_1\alpha_1 -
Q_2(\alpha_2-1) ) \Bigr] .
\eea
Define the numerator
\bea
N(\ell,p,\alpha) & = & \bigl[ G_4\bigl(p_1,p_2,\ell-
(Q_1\alpha_1+Q_2\alpha_2)\bigr) \\ \nonumber
& & G_4\bigl(p_3,p_4, \ell-(Q_1 (\alpha_1-1) + Q_2 \alpha_2)\bigr) \\
\nonumber
& & G_4\bigl(p_5,p_6, \ell -Q_1\alpha_1 - Q_2(\alpha_2-1) \bigr) \bigr] .
\eea

Now, because the $G$s are analytic, we can deal with the $\ell$ dependence of
$G$ by writing a symbolic Taylor expansion
\bea
 N(\ell,p,\alpha) & = & e^{\ell{\partial\over\partial
q}}\Bigl[G_4\bigl(p_1,p_2,q\bigr) \\ \nonumber
& & G_4\bigl(p_3,p_4, q +Q_1 \bigr) \\ \nonumber
& & G_4\bigl(p_5,p_6,q+Q_2 \bigr)\Bigr]_{q=-(Q_1\alpha_1+Q_2\alpha_2)}
\eea

and we just have to do the dimensionally regularized integral
\be
\int{d^{4-\epsilon}\ell
\over(2\pi)^{4-\epsilon}\mu^{-3\epsilon}}\,
{e^{\ell{\partial\over\partial q}}
\over\bigl[\ell^2-A^2+i\epsilon\bigr]^3}
\label{dr3}
\ee
where $ A^2 = Q_1^2 \alpha_1 (\alpha_1-1) + Q_2^2
\alpha_2(\alpha_2-1) + 2 Q_1
\cdot Q_2 \alpha_1 \alpha_2 + m^2$.

Again, we do this by manipulating the exponential like a power series, because
the basic assumption is analyticity in the momenta. Because of symmetry, only
even terms in $\ell$ contribute, and doing the $\ell$ integral yields a sum
 for the pole pieces\footnote{Note also that the factor of $2 /
\epsilon$ gets compensated
by a
factor of $1/2$ from a $\Gamma$ function}:
\be
{i \over 8 \pi^2 \epsilon \mu^{-3 \epsilon}} \sum_{r=1}^{\infty} {1 \over 4^r
r! (r-1)!}
 (A^2)^{r-1} \,\left[\left({\partial\over\partial q}\right)^2\right]^r
\label{sixsum}
\ee
where we have retained only the $1/\epsilon$ pole and the associated $\mu$
dependence. We cannot yet convert this sum to an angular integral, because
it starts at $r=1$, which just reflects the fact that the leading term in
the Taylor expansion gives a convergent Feynman integral, with no $1 /
\epsilon$ pole piece. To put the contribution of the sum back into the $G$'s,
we attempt to massage it further:
\bea
 & & \sum_{r=1}^\infty {1 \over 4^r r! (r-1)!} (A^2)^{r-1}
\,\left[\left({\partial\over\partial q}\right)^2\right]^r  \nonumber \\
& = &  {\partial^2 \over \partial x^2} \left[ \sum_{r=1}^\infty {1 \over 4^r
r!
(r+1)!}(A^2)^{r-1} x^{r+1}
\,\left[\left({\partial\over\partial q}\right)^2\right]^r \right]_{x=1}
\nonumber
\\
& = &  {\partial^2 \over \partial x^2} \left[ \sum_{r=0}^\infty {1 \over 4^r
r! (r+1)!} \bigl({x \over A^2} \bigr) (A^2)^{r} x^{r}
\,\left[\left({\partial\over\partial q}\right)^2\right]^r - {x \over A^2}
\right]_{x=1}
\nonumber
\\
& = &  {\partial^2 \over \partial x^2} \left[ \sum_{r=0}^\infty {1 \over 4^r
r! (r+1)!} \bigl({x \over A^2} \bigr) (A^2)^{r} x^{r}
\,\left[\left({\partial\over\partial q}\right)^2\right]^r \right]_{x=1}
 \label{sixsumfin}
\eea
Note that the finite sum subtracted off to redefine the sum to start at zero
gets annihilated by the differential operator.

The terms in the last sum \eqr{sixsumfin} are exactly
of the form seen before in \eqr{average}, and it can be
written in terms of a four dimensional Euclidean angular integral over a
finite range:
\be
{\partial^2 \over \partial x^2} \left[
{x \over A^2} \int {d \Omega^4_e \over \Omega^4} e^{\sqrt{x}A e
\cdot { \partial \over \partial q}} \right]_{x=1}
\ee
which can be put back into the $G$'s by inverting the Taylor expansion.
The contribution to the $\beta$ function for the nonlocal
$\phi^6$
vertex arising from a graph with three $G_4$ vertices is thus
 \bea
\lefteqn { \beta_{{G_{6[2,2,2]}} (p_1,\ldots p_5)} =} \label{sixfin}
\\ \nonumber
& & {1 \over 8 \pi^2
} {\partial^2 \over \partial x^2} \Biggl[
 \int {d \Omega^4_e \over \Omega^4} \int d \alpha_1 \, d
\alpha_2 {x \over A^2} \\
\nonumber
& & G_4(p_1,p_2, \sqrt{x}A e - (Q_1 \alpha_1 + Q_2
\alpha_2)) \\
\nonumber
& & G_4(p_3,p_4, \sqrt{x }A e - (Q_1 \alpha_1 + Q_2
\alpha_2)+ Q_1) \\
\nonumber
& & G_4(p_5,p_6, \sqrt{x }A e - Q_1 (\alpha_1-1)
- Q_2( \alpha_2-1) ) \Biggr]_{x=1} \\ \nonumber
& & +{\rm cross\atop terms}
\eea
where $e$ is a Euclidean unit vector and the square root of $A^2$ is real for
Euclidean momenta, so we have written it as $A$.

Alternatively, we can eliminate reference to the extraneous parameter $x$ by
the following trick. Consider the sum of \eqr{sixsum} again:
\be
\sum_{r=1}^\infty {1 \over 4^r r! (r-1)!} (A^2)^{r-1}
\,\left[\left({\partial\over\partial q}\right)^2\right]^r.
\ee
With
\be
p \equiv r-1
\ee
we obtain
\bea
& & \sum_{p=0}^\infty {1 \over 4^{p+1} p! (p+1)!} (A^2)^{p}
\,\left[\left({\partial\over\partial q}\right)^2\right]^{p+1} \\
&=& {1 \over 4}\left({\partial\over\partial q}\right)^2
\left[\sum_{p=0}^\infty {1 \over 4^{p} p! (p+1)!} (A^2)^{p}
\,\left[\left({\partial\over\partial q}\right)^2\right]^{p} \right] \\
&=& {1 \over 4}\left({\partial\over\partial q}\right)^2
\left[ \int {d \Omega^4_e \over \Omega^4} e^{ A e\cdot {\partial
\over \partial q}} \right],
\eea
 and using \eqr{kangular} for four
dimensions, $\int d \Omega^4 = 2\pi^2$,
we get a more compact form for the nonlocal $\beta$
function:
 \bea
\lefteqn { \beta_{{G_{6[2,2,2]}} (p_1,\ldots p_5)} =}
 \label{sixfin2} \\ \nonumber
& & {1 \over 64 \pi^4 } \left({\partial\over\partial
q}\right)^2
 \int d \Omega^4_e \int d \alpha_1 \, d
\alpha_2 \Bigl[ G_4(p_1,p_2, A
 e +q) \\ \nonumber
& & G_4(p_3,p_4, A e + q + Q_1) \\ \nonumber
& & G_4(p_5,p_6, Ae +q+Q_2) \Bigr]_{q=- (Q_1
\alpha_1 + Q_2 \alpha_2)} \\ \nonumber
& & +{\rm cross\atop terms} ,
\eea
The last form is free of extraneous parameters.

We reiterate that the evaluation of the $2-2-2$ diagram above highlights two
important features which will be crucial for
the application of the method to the general case below. First, for graphs
with more than two internal lines, the integral of the leading (zeroth) term
in the formal Taylor expansion does not give a divergence (it is
power-counting convergent), so the infinite sum for the poles starts only at
some higher order:
for a type $n$ graph ($n>1$)
the first divergent contribution comes from the $(n-2)$th
term in the Taylor expansion, so the sum of poles starts at $n-2$.
 To rewrite this sum in terms of the angular integral
over a Euclidean unit vector, we can subtract off a finite
sum which vanishes under the action of the differential operator of order
$n-1$ in the dummy parameter $x$. In fact, we can do much better - the
parameter $x$ can be eliminated by writing the pole sum a power of
$\left[\left({\partial\over\partial q}\right)^2\right]$ acting on a
$2n-2$ dimensional Euclidean unit angular integral of $G$'s. Let us
explain why this makes sense: For one-loop graphs with one or two internal
lines, the four dimensional Feynman integral has a divergence which begins
with the zeroth order term in the Taylor expansion (and higher terms in the
Taylor expansion are even more divergent), so the sum of poles starts at
zero. However, for graphs with more than two internal lines, enough powers
of momentum in the numerator are required to cancel the denominator powers
- thus the pole sum starts at a higher order. {\bf This sum can be made to
start at the zeroth order by defining the loop integral in the appropriate
number of higher dimensions}. Hence the resummation of the pole terms for a
sum starting at zero gives a higher dimensional angular integral.

\subsection{\normalsize \bf $2m$-point function of type $n$ at one loop}

For the non-trivial general graph with $n \geq 2$, after combining
denominators and shifting the loop
momenta, we get\footnote{The overall factor of $(n-1)!$ comes from the
Feynman trick.} the integral
\be
(n-1)! \int_0^1 \prod_{j=1}^{n-1} d \alpha_j \int_\ell {d^d \ell \over ( 2
\pi)^d} {1 \over D} {\prod_{i=1}^n \mu^{\Delta(i)}
G^i_{v_i +2} ( \ldots , \ell - \sum_{s=1}^{n-1} Q_s
\alpha_s + Q_{i-1})}
\label{general}
\ee
where $v_i$ is the number of external lines emanating from
the $i$th vertex, $\alpha_j$ are the Feynman parameters,
\be
D = [\ell^2 - ( \sum_{s=1}^{n-1} Q_s \alpha_s)^2 + (\sum_{r=1}^{n-1} Q_r^2
\alpha_r) -m^2 ]^n
\ee
is the combined denominator, and
\be
\Delta(i) = { \epsilon \over 2} v_i
\ee
 carries the renormalization scale
 dependence. The ellipsis denote the dependence of each $G^i$ on the
external momenta and the momenta $Q_i$ are defined as in
\eqr{momdef}.\footnote{ By momentum conservation at the last vertex, we may
make the structure of the last vertex simpler by making the
replacement
\be
 G^n_{v_n+_2}(\ldots,\ell - \sum_{s=1}^{n-1} Q_s \alpha_s +Q_{n-1})
\rightarrow
 G^n_{v_n+_2}(\ldots,-\ell + \sum_{s=1}^{n-1} Q_s \alpha_s).
\label{repl}
\ee}

 For
the renormalization of a $2m$ point function with $n$ internal lines
 \be
\sum_{i=1}^{i=n} v_i = 2m \ee
so summing up the $\mu$ dependence from all the vertices gives the overall
power of $\mu$ appearing as $\mu^{\Delta_{m}}$ where
\be
\Delta_{m} \equiv \sum_{i=1}^{i=n} v_i( {\epsilon \over 2}) = m \epsilon .
\ee
Now, using again the fact that all the $G$'s are analytic, we can write
\bea
&&\prod_{i=1}^{n} G^i_{v_i+2}(\ldots,l-\sum_{s=1}^{n-1} Q_s\alpha_s + Q_{i-1})
\\ \nonumber
& = &e^{\ell {\partial \over \partial q}} \left[ \prod_{i=1}^{n}
G^i_{v_i+2}(\ldots,q+Q_{i-1}) \right]_{q= - \sum_{s=1}^{n-1} Q_s \alpha_s}
.\eea

Now we just have to do the dimensionally regularized integral
\be
\int{d^{4-\epsilon}\ell
\over(2\pi)^{4-\epsilon}\mu^{-\Delta_m \epsilon}}\,
{e^{\ell{\partial\over\partial q}}
\over\bigl[\ell^2-A^2+i\epsilon\bigr]^n}
\label{drf}
\ee
where
\be
A^2 = ( \sum_{s=1}^{n-1} Q_s \alpha_s)^2 - (\sum_{r=1}^{n-1} Q_r^2
\alpha_r) + m^2 .
\ee

Remembering the analyticity in momenta, we do this as usual by manipulating
the
exponential like a power series, with only even terms in $\ell$ contributing.
Doing the integral as before using the key formula \eqr{key}, and expanding
the
$\Gamma$ function using \eqr{gam}, with
\be
\Gamma(n-r-2+{\epsilon \over 2}) = {2 \over \epsilon} {(-1)^{2+r-n} (2+r-n)!}
\ee
for $n \leq r+2$,
 we get the pole piece
\be
(n-1)! {i \over 8 \pi^2 \mu^{-\Delta_m} \epsilon} \sum_{r=n-2}^{\infty}
{ (-1)^{2+r-n}{(-1)}^{n+r}
(A^2)^{2+r-n}
\over {(n-1)! 4^r r! (2+r-n)!}} \,\left[\left({\partial\over\partial
q}\right)^2\right]^r \label{mainsum}.
\ee
Quite nicely, the $(n-1)!$ coming from the integration formula cancels with
the
$(n-1)!$ appearing from the Feynman trick.
 The sum \eqr{mainsum} can then be
written as
\bea
& & { i \over 8 \pi^2 \mu^{-\Delta_m} \epsilon} \sum_{r=n-2}^{\infty}
 {(A^2)^{2+r-n}
\over { 4^r r! (2+r-n)!}} \left[\left({\partial\over\partial
q}\right)^2\right]^r \nonumber \\
& = &
{ i \over 8 \pi^2 \mu^{-\Delta_m} \epsilon}
{\partial^{n-1} \over \partial x^{n-1}}
 \left[ \sum_{r=n-2}^{\infty}
{(A^2)^{2+r-n} x^{r+1}
\over  { 4^r r! (r+1)!}} \left[\left({\partial\over\partial
q}\right)^2\right]^r \right]_{x=1} \nonumber \\
& = &
{ i \over 8 \pi^2 \mu^{-\Delta_m} \epsilon}
{\partial^{n-1} \over \partial x^{n-1}}
 \left[{x \over {(A^2)}^{n-2} } \sum_{r=0}^{\infty}
{(A^2)^{r} x^r
\over  { 4^r r! (r+1)!}} \left[\left({\partial\over\partial
q}\right)^2\right]^r -T \right]_{x=1}
\eea
where we have defined the finite sum
\be
T = {x \over {(A^2)}^{n-2} }
\sum_{r=0}^{n-3} {(A^2)^{r} x^r
\over  { 4^r r! (r+1)!}} \left[\left({\partial\over\partial
q}\right)^2\right]^r .
\label{tsum}
\ee
taken to be vanishing for $n<3$. As in the last example, this finite sum is
subtracted off to make the first sum start at $r=0$. Now the differential
operator is of order $n-1$, and the finite sum is of order $n-2$, so the
finite sum gets annihilated by the differential operator, giving finally the
pole piece:
\be
{ i \over 8 \pi^2 \mu^{-\Delta_m} \epsilon}
{\partial^{n-1} \over \partial x^{n-1}}
 \left[{x \over {(A^2)}^{n-2} } \sum_{r=0}^{\infty}
{(A^2)^{r} x^r
\over  { 4^r r! (r+1)!}} \left[\left({\partial\over\partial
q}\right)^2\right]^r \right]_{x=1}.
\ee
Now, inverting as usual using \eqr{average} for a Euclidean unit vector, we
may
cast this pole contribution as an integral over a finite
four-dimensional Euclidean angular region:
\bea
\lefteqn {{\rm Pole} \, = {i\over 8 \pi^2 \epsilon \mu^{-m\epsilon} }
{\partial^{n-1} \over \partial x^{n-1}} } \label{finalom} \\
\nonumber
& & \Biggl[  \int {d \Omega^4_e \over \Omega^4}
\int_0^1 \prod_{j=1}^{j=n-1} d \alpha_j
{x \over (A^2)^{n-2}} \\ \nonumber
& & \prod_{i=1}^{i=n} G^i_{v_i+2} \left(\ldots,
\sqrt{x} A e - (\sum_{s=1}^{n-1} Q_s \alpha_s) + Q_{i-1} \right) \Biggr]_{x =
1}
\eea
which gives the $\beta$ function in the form\footnote{With the property that
at the last vertex in the product
we may make the replacement, by momentum conservation,
\bea
\lefteqn { G^n_{v_n+2} (\ldots, \sqrt{x} A e - (\sum_{s=1}^{n-1} Q_s
\alpha_s) + Q_{n-1} )}
\nonumber \\
& \equiv & G^n_{v_n+2} (\ldots,
 \sqrt{x} A e + (\sum_{s=1}^{n-1} Q_s \alpha_s) ).
\label{finaldef}
\eea}
\bea
\lefteqn {\beta_{{G_{2m[v_1,\ldots,v_n]}} (p_1,\ldots,p_{2m-1})} =}
\label{finaleq} \\
\nonumber
& & { 1 \over 16 \pi^4
 } { \partial^{n-1} \over \partial x^{n-1} }  \left[ \int d \Omega^4_e
\int_0^1 \prod_{j=1}^{j=n-1} d \alpha_j
{x \over (A^2)^{n-2}} \right. \\ \nonumber
& & \left. \prod_{i=1}^{i=n} G^i_{v_i+2} \left(\ldots,
 \sqrt{x} A e - (\sum_{s=1}^{n-1} Q_s \alpha_s) + Q_{i-1} \right) \right]_{x =
1} \\ \nonumber
& & +{\rm cross\atop terms},
\eea
However, we can again do much better, and obtain a more compact expression
in terms of a higher dimensional angular integral. Referring back to
\eqr{mainsum}, we can write
\be
 \sum_{r=n-2}^{\infty}
{ (A^2)^{2+r-n}
\over { 4^r r! (2+r-n)!}} \,\left[\left({\partial\over\partial
q}\right)^2\right]^r
\ee
in terms of
\be
p \equiv r-n+2,
\ee
for
\be
n \geq 3,
\ee
as
\bea
& & \sum_{p=0}^{\infty}
{ (A^2)^p
\over { 4^{p+n-2} p! (p+n-2)!}} \,\left[\left({\partial\over\partial
q}\right)^2\right]^{p+n-2} \\
& =& {1 \over 4^{n-2}} \left[\left({\partial\over\partial
q}\right)^2\right]^{n-2} \left[ \sum_{p=0}^{\infty}
{ (A^2)^p \left[\left({\partial\over\partial
q}\right)^2\right]^{p}
\over { 4^{p} p! (p+n-2)!}} \, \right] \nonumber \\
&=& {1 \over 4^{n-2}} \left[\left({\partial\over\partial
q}\right)^2\right]^{n-2} \left[ \int {d
\Omega^{2n-2}_e \over \Omega^{2n-2}} e^{A e \cdot {\partial \over
\partial q}} \right]
\eea
so that we finally get the pole
\bea
\lefteqn {{\rm Pole} \, = {i\over 2^{2n-1} \pi^2 \epsilon \mu^{-m\epsilon} }
\left[\left({\partial\over\partial q}\right)^2\right]^{n-2} } \nonumber \\
& &\Biggl[ \int {d \Omega^{2n-2}_e \over \Omega^{2n-2}}
\int_0^1 \prod_{j=1}^{j=n-1} d \alpha_j \\\nonumber
& & \prod_{i=1}^{i=n} G^i_{v_i+2}
\left(\ldots, A e +q
 + Q_{i-1} \right)  \Biggr]_{q=-
(\sum_{s=1}^{n-1} Q_s \alpha_s)}
\eea
which, using \eqr{kangular} yields the contribution to the $\beta$ function
in terms of a $2n-2$ dimensional angular integral over a finite Euclidean
region:
\bea \lefteqn {\beta_{{G_{2m[v_1,\ldots,v_n]}}
(p_1,\ldots,p_{2m-1})} =} \label{finaleq2} \\
\nonumber
& & { (n-2)! \over 4^n \pi^{n+1} } \left[ \left({\partial\over\partial
q}\right)^2\right]^{n-2} \Biggl[ \int d \Omega^{2n-2}_e
\int_0^1 \prod_{j=1}^{j=n-1} d \alpha_j \\ \nonumber
& &  \prod_{i=1}^{i=n} G^i_{v_i+2} \left(\ldots,
 A e +q + Q_{i-1} \right) \Biggr]_{q=- (\sum_{s=1}^{n-1} Q_s \alpha_s)} \\
\nonumber
& & +{\rm cross\atop terms}.
\eea

Expressions \eqr{finaleq}, \eqr{finaldef} and \eqr{finaleq2} for the $\beta$
function of an arbitrary nonlocal coupling are the most general results of
this paper.\footnote{It is verified easily that with $n=2, m=2$ the $\beta$
function calculation for the nonlocal $\phi^4$ coupling, as in section 1, is
reproduced.
For $n=3, m=3$, and $n=1, m =2$, the previous
results for the $\phi^6$ maximally convergent graph ($2-2-2$) and
$\phi^4$ tadpole are also verified from the general expression.}

\section{Concluding Remarks}

We have illustrated explicitly how to obtain
the renormalization group coefficients in a nonlocal scalar
field theory. Note that our general results were computed with massive fields,
and for this reason techniques such as the Gegenbauer polynomial method of
\cite{CKT},
valid only for massless propagators, are not useful here.

We regard our results as suggestive, but not final. We have obtained
integro-differential renormalization group equations by the brute-force
technique of exanding,
renormalizing and then resumming. We would like to find the rule that allows
us to do this in one step, rather than three.  It would be interesting if, by
modifying the analytic structure of the nonlocal theory, we could actually
give a one-step prescription to isolate the divergent part of the Feynman
integral, without ever expanding in terms of a formal Taylor expansion. Work
along these lines is in progress \cite{VBHG}.

We would like to thank Peter Cho, Mike Dugan and Ben Grinstein for numerous
discussions.

\end{document}